\documentclass{iau}
\usepackage{graphicx}

\title[Evolution of galaxy sizes] %% give here short title %%
{The evolution of galaxy sizes}

\author[]   %% give here short author list %%
{Bianca M. Poggianti$^1$,
Rosa Calvi$^2$, Daniele Bindoni$^2$,  Mauro D'Onofrio$^2$,  
Alessia Moretti$^2$, Tiziano Valentinuzzi,  
Giovanni Fasano$^1$, Jacopo Fritz$^3$,  
Gabriella De Lucia$^4$, Benedetta Vulcani$^2$, Daniela Bettoni$^1$, Marco
Gullieuszik$^1$ \and Alessandro Omizzolo$^1$}

\affiliation{$^1$ INAF-Astronomical Observatory of Padova, vicolo dell'Osservatorio 5, 35122 Padova, Italy \\
email: {\tt bianca.poggianti@oapd.inaf.it} \\[\affilskip]
$^2$ Department of Astronomy, University of Padova \\[\affilskip]
$^3$ Sterrenkundig Observatorium, University of Gent, Belgium 
\\[\affilskip] $^4$ INAF-Astronomical Observatory of Trieste
}
\pubyear{2013}
\volume{295}  %% insert here IAU Symposium No.
\pagerange{xx--xx}
\jname{The intriguing life of massive galaxies}
\editors{D. Thomas, A. Pasquali \& I. Ferreras, eds.}
\begin{document}

\maketitle

\begin{abstract}
We present a study of galaxy sizes in the local Universe as a function
of galaxy environment, comparing clusters and the general field.
Galaxies with radii and masses comparable to high-z massive and
compact galaxies represent 4.4\% of all galaxies more massive than $3
\times 10^{10} M_{\odot}$ in the field. Such galaxies are 3 times more
frequent in clusters than in the field.  Most of them are early-type
galaxies with intermediate to old stellar populations. There is a
trend of smaller radii for older luminosity-weighted ages at fixed
galaxy mass. 
%This effect is much more pronounced in clusters than in the field. 
We show the relation between size and luminosity-weighted
age for galaxies of different stellar masses and in different
environments.  We compare with high-z data to quantify the evolution
of galaxy sizes.  We find that, once the progenitor bias due to the
relation between galaxy size and stellar age is removed, the average
amount of size evolution of individual galaxies between high- and
low-z is mild, of the order of a factor 1.6.
\end{abstract}

\firstsection % if your document starts with a section,
\section{Introduction}

Many recent observational studies have found a population of high-z
($z=1-2.5$), massive ($M_{\star}> 10^{10} M_{\odot}$), compact ($R_e <
2 \rm kpc$) galaxies (e.g. Daddi et al. 2005, Trujillo et al. 2006,
Cimatti et al. 2008, van Dokkum et al. 2008, Saracco et al. 2009,
Cassata et al. 2011 and others). In the great majority 
of these studies, galaxies are selected
to be already passively evolving (devoid of ongoing star formation) at
the redshift they are observed. Passive galaxies at high-z have been
found to display a wide range in sizes, from extremely compact
to those whose sizes are comparable to normal galaxies in the local
Universe (Saracco et al. 2009, Mancini et al. 2010, Cassata et al. 2011).

Most high-z works use as local comparison the {\it r}-band Sloan
median mass-size relation
for galaxies with Sersic index $n \geq 2.5$ 
from Shen et al. (2003).
Trujillo et al. (2009) and Taylor et al. (2010) searched for massive compact 
galaxies in Sloan and both found very few such galaxies in the general field
at low redshift.

Here we present a search for massive and compact
galaxies in different environments at low redshift using two galaxy
samples: for the general field we use the Padova Millennium Galaxy
and Group Catalogue at $z=0.04-0.1$ (PM2GC, Calvi et al. 2011), and for
galaxy clusters the WIde-field Nearby Galaxy cluster Survey at
$z=0.04-0.07$ (WINGS, Fasano et al. 2006).

``Superdense'' galaxies, with galaxy stellar masses and effective sizes
similar to those of high-z galaxies 
($\Sigma_{50} = (0.5 * M_{\star})/(\pi * 
{R_e}^2) > 3 \times 10^9 M_{\odot} \rm kpc^{-2}$), were found to represent
$\sim 20\%$ of all galaxies with $3 \times 10^{10} < M_{\star} < 4 \times
10^{11} M_{\odot}$ 
in WINGS clusters based on V-band images
(Valentinuzzi et al. 2010, hereafter V10). 
Given the paucity
of similar galaxies in the field studies mentioned above, this suggested
a strong environmental dependence in the frequency of massive
and compact galaxies at low-z (V10).

V10, and a number of other works (Saracco et al. 2011, Cassata et al. 2011,
Cappellari et al. 2012 and these proceedings, to name a few) also
found that galaxy sizes depend on the age of stars: galaxies with
older luminosity-weighted ages (hereafter, LW ages) are smaller.
As a consequence, selecting the oldest galaxies at high-z (those
that have already stopped forming stars at $z=1-2.5$) means selecting
the most compact ones. At lower redshifts, larger galaxies stop
forming stars and enter the passive galaxy samples, causing
the median mass-size relation to move to larger radii at fixed mass.
Hence, in order to measure the correct amount of size
evolution, the sizes of high-z galaxies should be compared only with those
of galaxies whose LW age is so old that they were already  passive
at high-z (V10).

Moreover, in clusters at $z=0.4-0.8$, superdense galaxies are 40\% of the
galaxy population, and the mass-size relation in clusters
evolves very little between $z\sim 0.6$ and $z=0$ (Valentinuzzi
et al. 2010b).

\section{Results in the field}

Recently (Poggianti et al. 2012, submitted), we have searched for
massive compact galaxies in the field
using the PM2GC, a sample of low-z galaxies representative of the
general field population drawn from the Millennium Galaxy Catalogue
(MGC, Liske et al. 2003, Driver et al. 2005). We measured galaxy radii and
galaxy morphologies with GASPHOT (Pignatelli \& Fasano 2006) and
MORPHOT (Fasano et al. 2012), respectively, from the MGC B-band images.  
Size and stellar mass estimates are in very good agreement 
with other existing measurements.

After inspecting each superdense candidate, we found 44 superdense
galaxies out of 995 galaxies in the PM2GC, corresponding to a fraction of
4.4\%$\pm 0.7$ among galaxies with $M_{\star} > 3 \times 10^{10}
M_{\odot}$.  The corresponding fraction in clusters from B-band WINGS
images is 11.8\%$\pm1.7$ (about half of the V-band
percentage found by V10). This is due to the small but systematic
change of galaxy radii with wavelength, with radii increasing
using bluer passbands.

Most of the PM2GC superdense galaxies are red, early-type galaxies (70\% S0s
and 23\% ellipticals), quite flattened, with a mean
Sersic index $n=2.8$, and have intermediate-to-old stellar 
populations, with a mean LW age =5.5 Gyr and a mean mass-weighted
age =9.3 Gyr.

\begin{figure}[b]
\vspace*{-2.0 cm}
\begin{center}
 \includegraphics[width=4.5in]{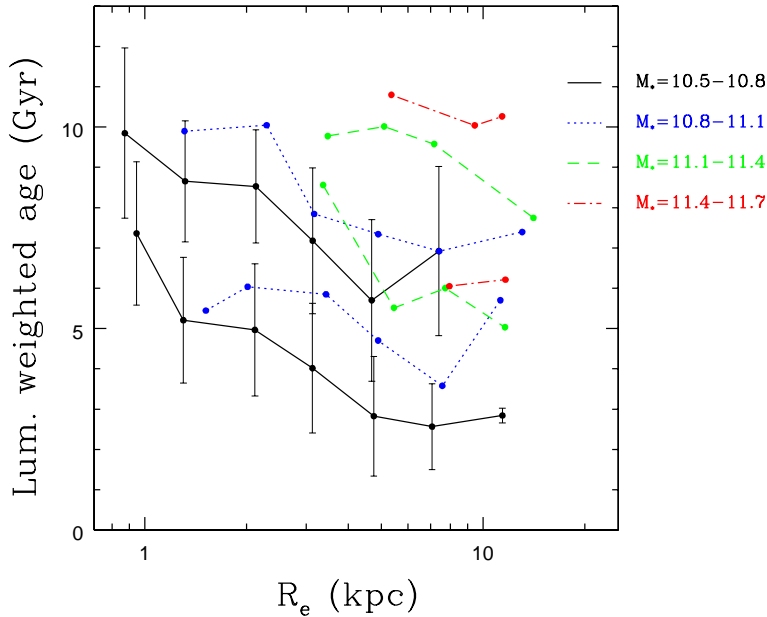} 
% \vspace*{-1.0 cm}
 \caption{Median LW age of galaxies as a function of effective radius
for four bins of galaxy masses (see legend). For each mass bin, the lowest line
refers to the field and the upper line to clusters.
}
   \label{fig1}
\end{center}
\end{figure}

Figure~1 illustrates the relation between size and LW age
for galaxies in different mass bins and different environments
(for each galaxy mass, the lower line is the PM2GC, and the upper line
is WINGS).
At given size and environment, 
more massive galaxies have older ages, both in clusters
and in the field, following downsizing. The plot shows
that, at given mass and environment, galaxies with smaller radii (more compact)
are older. This effect is present both in clusters and in
the field, although our analysis (not shown) finds that
 the dependence of the median mass-size
relation on stellar age is stronger in clusters than in the field.

As a consequence, the population of passive galaxies at low-z includes
both the old, on average small galaxies and those galaxies that have stopped
forming stars at later times that have on average larger sizes.
This needs to be taken into account when trying to quantify
the evolution of the high-z samples.

\subsection{Size evolution: comparison with high-z studies}

\begin{figure}[b]
\vspace*{-4.0 cm}
\begin{center}
 \includegraphics[width=5.0in]{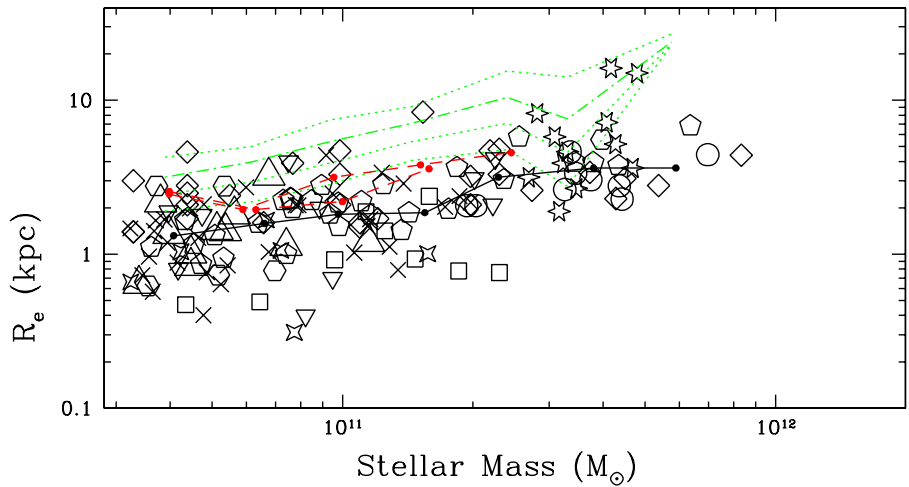} 
% \vspace*{-1.0 cm}
 \caption{Median mass-size relation (solid line) of
high-z galaxies (large symbols) compared to the
relation of PM2GC and WINGS galaxies with LW ages $\geq 10$ Gyr
(dotted) and of
all PM2GC galaxies (the dashed-dotted line is the median, dotted
lines are 1 and 2$\sigma$).
}
   \label{fig1}
\end{center}
\end{figure}

Figure~2 shows a compilation of various datasets of galaxies at $z\sim 1-2.5$
(different symbols, Daddi et al. 2005, Trujillo et al. 2006, van Dokkum et 
al. 2008, Cimatti et al. 2008, van der Wel et al. 2008, Saracco et al. 2009, 
Damjanov et al. 2009, Mancini et al. 2010, Cassata
et al. 2011) and their median mass-size relation (solid line).

To quantify the evolution in size, we compare this relation
with the median mass-size relation of {\it old} (LW age $\geq 10$ Gyr)
galaxies in WINGS and the PM2GC (two dotted lines in Fig.~2). We find
an average evolution of 0.2-0.25dex, a factor 1.6-1.8, using WINGS
and PM2GC, respectively. This is half of what is found
using all, passive or $n \geq 2.5$ galaxies in the PM2GC, as
usually done in the literature. The evolution of galaxy sizes is therefore
much smaller than would be (erroneously) inferred at face value considering
as local descendants the whole passive galaxy population regardless
of age.

Using old WINGS or old PM2GC galaxies does not significantly change the 
results. It is however important to stress that our simulations
(Millennium Simulation + semi-analytic model presented in De Lucia
\& Blaizot 2007) predict that 60\% of all galaxies that at $z \sim 2$
are massive ($\geq 10^{11} M_{\odot}$) and passive end up in 
haloes with masses above $10^{14} M_{\odot}$ by $z=0$, therefore
in WINGS-like clusters in the local Universe. Hence, 
a large fraction of the high-z passive and massive galaxies 
have evolved into today's cluster galaxies. This may explain the strong
environmental dependence of the superdense incidence we find at low-z,
and indicates that galaxies in nearby clusters are the most appropriate
local counterparts to the high-z studies.

%The results presented above can be found in Poggianti et al.
%(2012, submitted).

\end{document}